\documentclass[pre,showpacs,twocolumn, superscriptaddress,amsmath,amssymb]{revtex4}
\usepackage{amssymb}
\usepackage{amsmath}
\usepackage{graphicx}
\begin{document}
 \title{Squared-field
amplitude modulus and radiation intensity nonequivalence within
nonlinear slabs}

\author{Alberto Lencina}

\affiliation{Departamento de F\'isica, Centro de Ci\^encias Exatas
e da Natureza, Universidade Federal da Para\'iba, Caixa Postal
5008 CEP 58051-970, Jo\~ao Pessoa, PB Brazil}

\email[Corresponding author A. L.: ] {agl@fisica.ufpb.br}

\author{Pablo Vaveliuk}

\affiliation{Departamento de F\'isica, Universidade Estadual de
Feira de Santana, CEP 44031-460, Feira de Santana, BA Brazil}

\begin{abstract}
This paper presents a novel approach to wave propagation inside
the Fabry-Perot framework. It states that the time-averaged
Poynting vector modulus could be nonequivalent with the
squared-field amplitude modulus. This fact permits the
introduction of a new kind of nonlinear medium whose nonlinearity
is proportional to the time-averaged Poynting vector modulus. Its
transmittance is calculated and found to differ with that obtained
for the Kerr medium, whose nonlinearity is proportional to the
squared-field amplitude modulus. The latter emphasizes the
nonequivalence of these magnitudes. A space-time symmetry analysis
shows that the Poynting nonlinearity should be only possible in
noncentrosymmetric materials.
\end{abstract}


\date{\today}
\maketitle

\section{Introduction}
A classical topic in electromagnetism is the study of wave
transmission in a finite parallel-plane faces medium, know as
Fabry-Perot resonator. When the medium presents nonlinear
behavior, bistability appears \cite{gibbs}. To explain this
phenomenon, the nonlinear Fabry-Perot resonator (NLFP) was
modelled by a third order susceptibility or Kerr-type nonlinearity
\cite{marburger}. At monochromatic plane wave excitation, the NLFP
stationary regime is summarized in a non-time-dependent nonlinear
wave equation for the complex field amplitudes, the nonlinear
Helmholtz equation (NLHE). Hence, the reflectance and
transmittance problem reduces to finding the NLHE solution with
appropriate boundary conditions for the field amplitude modulus
and phase.

The NLHE complexity led to approximate methods of resolution. Many
approaches consider two counter-propagating waves in the medium
and the analysis is done by separately considering the effects on
each wave \cite{marburger,miller,danckaert,biran}. Unfortunately,
the linear superposition principle is no longer valid in nonlinear
media and the separation of the electromagnetic field in these
back and forth waves is meaningless. As a result, the NLHE
separation into two equations, one for each wave, is only possible
by neglecting various coupling nonlinear terms that would give an
important contribution to the accuracy of the transmittance
results. Moreover the Slowly Varying Envelope Approximation (SVEA) is
often applied to these waves \cite{marburger,miller,danckaert}
when, yet within the counter-propagating wave approach, its
validity was questioned \cite{biran}. Also, the boundary
conditions were simplified rather than rigorously treating them
\cite{marburger,fobelets}. The above facts suggest that all these
approximated approaches could not physically be equivalents to the
exact problem.

The work done by Chen and Mills exactly solve the NLFP for an
absorptionless Kerr-type medium \cite{mills1}. The proper of its
resolution method was to assume a general complex field within the
medium, disregarding the concept of counter-propagating waves.
Chen and Mills derive a two coupled equation system for the field
amplitude modulus and phase together with general boundary
conditions, thus obtaining an analytic-transcendental solution for
the transmittance of the NLFP.

On the other hand, their work permitted us to note an implicit
difference between the time-averaged Poynting vector modulus, i.e.
the electromagnetic radiation intensity $I$, and the squared-field
amplitude modulus $(|E|^{2})$ inside the nonlinear medium. If the
nonequivalence of these magnitudes were true, it could change
certain well-established fundamental concepts in classical
electrodynamics. This fact motivated us to develop a novel
approach to wave propagation in nonlinear media inside the
Fabry-Perot framework called \textit{S-Formalism}. It introduces a
new variable related to the time-averaged Poynting vector which
states that its magnitude could be nonequivalent with the
squared-field amplitude modulus contrary to commonly accepted.
Furthermore, the S-Formalism presents two important advantages: it
permits to directly monitor the radiation intensity within the
medium, and it avoids approximations, such as the SVEA,
simplification of boundary conditions, and so on.

The fact that the time-averaged Poynting vector modulus be
nonequivalent with the squared-field amplitude modulus, as the
S-Formalism will show, implies that the nonlinearity of Kerr-type
media is not proportional to the intensity which is contrary to
what has been established to-date. This assertion leads to the
following question regarding the modelling of the NLFP: is it a
Kerr-type nonlinearity, or does it vary proportionately to the
intensity? As this question does not have a definitive answer, the
existence of the latter cannot be denied. Then, we define the
\textit{Poynting medium} as the medium where nonlinearity is
proportional to intensity. Thus, our objective is to solve the
\textit{Poynting-NLFP} through the S-Formalism comparing the
resultant transmittance with that obtained for a
\textit{Kerr-NLFP} to remark the nonequivalence between
squared-field amplitude modulus and radiation intensity.
\begin{figure}[t]
\includegraphics[width=6.5cm,height=4.5cm]{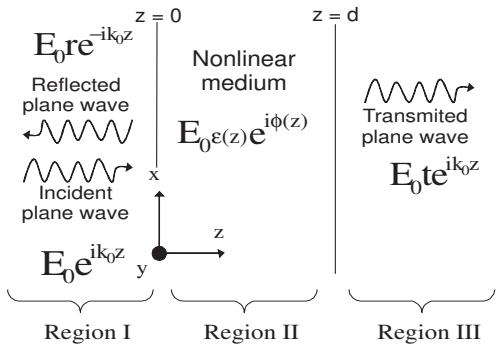}
\caption{An harmonic plane wave strikes a nonlinear Fabry-Perot
resonator, to be reflected and transmitted. The Regions I and III
constitute, for simplicity, the same linear dielectric medium
(e.g., air).\\[-20pt]} \label{fig=problema}
\end{figure}

In section II, we derive the S-Formalism in the following form:
first the time-averaged Poynting vector assuming harmonic fields
is calculated. Then, a dimensionless variable which is
proportional to intensity is introduced. Thus, a set of field
evolution differential equations and the general boundary
condition equations on the new field variable are obtained, which
constitutes the S-Formalism. In section III, the new approach is
applied to derive the transmittance for the Poynting-NLFP and the
results compared with that obtained for a Kerr-NLFP. A brief
discussion about the possibily of existence and observation of
Poynting media is given. Finally, in section IV, we conclude.

\section{The S-Formalism approach}
Refering to Figure \ref{fig=problema}, we start writing the
linearly polarized transversal harmonic electromagnetic fields of
frequency $\omega$ as
\begin{subequations}
\begin{eqnarray}
\mathbf{E}_{\ell}(z,t)=\frac{1}{2}\left(E^{\omega}_{\ell}(z)e^{-i\omega
t}+c.c.
\right)\mathbf{\hat{i}},\\[3pt]
\mathbf{H}_{\ell}(z,t)=\frac{1}{2}\left(H^{\omega}_{\ell}(z)e^{-i\omega
t}+c.c. \right)\mathbf{\hat{j}},
\end{eqnarray}
\end{subequations}
where $E^{\omega}_{\ell}(z)$ and $H^{\omega}_{\ell}(z)$ are the
non-time-dependent complex amplitudes for $\ell= $I, II, III. From
region I, a plane wave of amplitude $E_{0}$ and wave vector
$k_{0}$ impinges perpendicularly on a nonmagnetic, isotropic and
spatially nondispersive medium of thickness $d$ (region II). The
optical field is assumed to maintain its polarization along this
region so that a scalar approach is valid. The reflected and
transmitted plane waves have amplitudes $rE_{0}$ and $tE_{0}$ with
$r$ and $t$ the complex reflection and transmission coefficients,
respectively. Then, at region I and III the spatial-dependent
complex amplitudes are given by
\begin{eqnarray}
E_{I}^{\omega}(z)&=&E_{0}\left(e^{i k_{0}z}+r e^{-i k_{0}z}
\right)\label{eq=fielI},\\[3pt]
E_{III}^{\omega}(z)&=&E_{0}t e^{i k_{0}z}\label{eq=fielIII}.
\end{eqnarray}
Similarly to Ref. \cite{mills1}, at region II, we write down the
following \textit{ansatz} for the spatial-dependent complex
amplitude of the electric field:
\begin{equation}
E^{\omega}_{II}(z)\:=\:E_0 \mathcal{E}(z) e^{i
\phi(z)},\label{eq=camporegII}
\end{equation}
where the dimensionless amplitude $\mathcal{E}(z)$, and phase
$\phi(z)$, are both real functions of $z$.

The time-averaged Poynting vector
$\langle\:\mathbf{E}_{\ell}(z,t)\times\mathbf{H}_{\ell}(z,t)\:\rangle$
can be easily calculated with the aid of Faraday's law, giving
\begin{eqnarray}
\langle\mathbf{S}\rangle_{\ell}=\frac{1}{2 \mu_{0}\omega}Im
\left\{\left[E^{\omega}_{\ell}(z)\right]^{*}\frac{\partial
E^{\omega}_{\ell}(z)}{\partial z}\right\}\mathbf{\hat{k}},
\end{eqnarray}
where $\mu_{0}$ is the vacuum permeability. From this expression
we calculate the intensity for the three regions:
\begin{subequations}\label{poyting}
\begin{eqnarray}
\langle S\rangle_{I}&=&I_{0}\left(1-|r|^{2}\right),\\[3pt]
 \langle S\rangle_{II}
&=&I_{0}\:k_{0}^{-1}
\mathcal{E}^{2}(z)\frac{\partial\phi(z)}{\partial
z} \equiv I_{0}S(z)\label{eq=defS}\label{eq=SregII},\\[3pt]
\langle S\rangle_{III}&=&I_{0}|t|^{2},\label{transmitancia}
\end{eqnarray}
\end{subequations}
where $I_{0}={k_{0}E^{2}_{0}}/({2 \mu_{0}\omega})$ is the incident
intensity. In region II, Eq. (\ref{eq=defS}) defines the
dimensionless field variable
\begin{equation}\label{S}
S\equiv k_{0}^{-1} \mathcal{E}^{2}\frac{\partial\phi}{\partial z},
\end{equation}
directly related with the intensity inside the medium and which
will characterize the S-Formalism. From Eq. (\ref{S}) it is clear
that if $\phi$ is not a linear function on $z$, as is often
happens in nonlinear media, then $S$ and $\mathcal{E}^{2}$ are
nonequivalents.

The next step is to derive the NLHE in terms of the classical
field variables ($\mathcal{E},\phi$), and transform it into a set
of equivalent equations in terms of ($\mathcal{E},S$). The NLHE is derived from the macroscopic Maxwell equations complemented by
appropriate constitutive relations. We assume that the
polarization $\mathbf{P}$, and current density $\mathbf{J}$, vary
only in the electric field direction with frequency $\omega$,
neglecting higher harmonics, and their spatial-dependent complex
amplitudes satisfy the following constitutive relations:
\begin{eqnarray}
P^{\omega}_{II}(z)&=&\epsilon_{0}\chi_{gen}[z,E^{\omega}_{II},H^{\omega}_{II}]\:
E^{\omega}_{II}(z),\\[3pt]
J^{\omega}_{II}(z)&=&\sigma_{gen}
[z,E^{\omega}_{II},H^{\omega}_{II}]\:E^{\omega} _{II} (z),
\end{eqnarray}
where $\epsilon_0$ is the vacuum permittivity and, $\chi_{gen}$
and $\sigma_{gen}$ are the generalized susceptibility and
conductivity respectively, that are real and contain the linear as
well as a possible nonlinear medium response. Note that the
constitutive relations are not explicitly written since cases
could exist where it is not possible to describe the nonlinear
polarization and current density by the classical electric field
power expansion.
Thereby, the scalar NLHE is
\begin{eqnarray}
 \Big[\frac{d^2}{dz^2}
+k_{0}^2(1+\chi_{gen}) +i \omega \mu_{0}\:\sigma_{gen}\Big]
E^{\omega}_{II}(z)=0. \label{eq=nlhe}
\end{eqnarray}
This equation constitutes the starting point to study several
linear and nonlinear monochromatic wave propagation phenomena
within the Fabry-Perot framework. Substituting Eq.
(\ref{eq=camporegII}) into (\ref{eq=nlhe}) and by using Eq.
(\ref{S}), we derive the following set of spatial evolution
equations for the field variables $\mathcal{E}(z)$ and $S(z)$:
\begin{subequations}
\begin{eqnarray}
&&\frac{d^2\mathcal{E}}{d z^2}
+k_{0}^{2}\left((1+\chi_{gen}[z,\mathcal{E},S])\:\mathcal{E}
-\frac{S^{2}}{\mathcal{E}^{3}}\right)=0,
\\[3pt]
&&\frac{d S}{d z}+\frac{\omega}{k_{0}}\mu_{0}\:\sigma_{gen}
[z,\mathcal{E},S]\:\mathcal{E}^{2}\label{eq=poyntinhtheo}=0.
\end{eqnarray}
\label{eq=formalismoS}
\end{subequations}
To guarantee the physical content of the solution, these equations
must be necessarily complemented with the following boundary
conditions: the continuity of the tangential components of the
electric and magnetic field at the interfaces. The general
boundary conditions were rigourously derived in Ref.
\cite{mills1}: four equations as functions of $(\mathcal{E},\phi)$
at $z=0$ and $z=d$ which, by using Eq. (\ref{S}), are transformed
into three equations in terms of the variables $(\mathcal{E},S)$
to give
\begin{subequations}
\begin{eqnarray}
&&\left(\mathcal{E}(0)+\frac{S(0)}{\mathcal{E}(0)} \right)^{2}+
\left(\frac{1}{k_{0}} \frac{d\mathcal{E}}{dz} \Big|_{z=0} \right)^{2}=4,\\[3pt]
&&S(d)-\mathcal{E}^{2}(d)=0,\\[3pt]
&&\frac{d\mathcal{E}}{dz}\Big|_{z=d}=0.
\end{eqnarray}
\label{eq=condecont}
\end{subequations}
From Eqs. (\ref{poyting}) and (\ref{eq=condecont}), the
transmittance is obtained as
\begin{equation}
T=|t|^2=S(d),
\end{equation}
and the energy conservation is guaranteed thorough the expression:
\begin{equation}
|r|^2+|t|^2=1-\left(S(0)-S(d)\right).
\end{equation}
This equation establishes that the reflectance and transmittance
are limited by the boundary values of the time-averaged Poynting
vector. For a nonabsorbent medium $S(d)=S(0)$, then
$|r|^{2}+|t|^{2}=1$.

Equations (\ref{eq=formalismoS}) and (\ref{eq=condecont})
represent the "S-Formalism" which were derived without assuming
approximations such as the counter-propagating waves, SVEA,
simplifications on the boundary conditions, and so on. Also, note
that Eq. (\ref{eq=poyntinhtheo}) represents the time-averaged
Poynting Theorem applied to the problem of harmonic fields
simplifying the interpretation of $\sigma_{gen}$ as the
dissipation properties of the medium. In particular, when
$\sigma_{gen}=0$, the dimensionless intensity $S$ is a constant
fixed by the boundary conditions. Furthermore, through $S(z)$ it
is possible to monitor directly the intensity along the medium as
a function of the spatial coordinate, as opposed to using the
conventional formalism.

The S-Formalism is useful to analyze the linear case as well as
the nonlinear one. Before to study the latter, i.e. the comparison
between the Poynting and Kerr media in an effort to show the
explicit difference between $S$ (or $I$) and $\mathcal{E}^2$ (or
$|E^{2}|$) in nonlinear media, we refer to the linear case.
According to our analysis \cite{ajp}, there are only two
situations where the relationship $I=cte |E^2|$ holds true:
firstly, a single plane wave propagates in a infinite or
semi-infinite linear dielectric characterized by $\sigma_{gen}=0$
and $\chi_{gen}=\chi^{(1)}$ where $\chi^{(1)}$ is the linear
susceptibility. Under these conditions, Eqs.
(\ref{eq=formalismoS}) relate the constants $S$ and $\mathcal{E}$
by $S=(1+\chi^{(1)})^{1/2}\:
 \mathcal{E}^2$; secondly, a single
plane wave propagates in a semi-infinite linear absorber
characterized by $\chi_{gen}=\chi^{(1)}$ and $\sigma_{gen}=\sigma$
where $\sigma$ is the ohmic conductivity and such that
$S(z)\propto \mathcal{E}^2(z)$, being both proportional to a
decreasing exponential function of $z$. On the contrary, when the
medium is finite, e.g. a Fabry-Perot with boundary conditions at
interfaces, $S$ is no longer equivalent to $\mathcal{E}^2$, not
even for the linear dielectric case because $S$ is a constant and
$\mathcal{E}^2$ is an oscillating function of $z$ \cite{ajp}.

\section{The Poynting medium}
\subsection{Constitutive relations and transmittance results}

At this point, we introduce the Poynting medium by the following
constitutive relations:
\begin{eqnarray}
\chi_{gen}&=&\chi^{(1)}+\gamma I_0 S(z),\label{constitutive}\\[3pt]
\sigma_{gen}&=&0,
\end{eqnarray}
where $\gamma$ is the nonlinear coefficient. Eqs.
(\ref{eq=formalismoS}) have a simple analytical solution given by
\begin{eqnarray}
S(z)&=&S_{0},\\[3pt]
\mathcal{E}(z)&=&\sqrt{\frac{S_{0}}{2}\left[
\left(1-\frac{k_{0}^{2}}{k_{1}^{2}}\right)\cos \left[2 k_{1}
(z-d)\right]+1+\frac{k_{0}^{2}}{k_{1}^{2}}\right]},\:\:\:\:\:\:\:
\label{eq=epsPoy}
\end{eqnarray}
where $\gamma>0$ and $k_{1}^{2}=k_{0}^{2}(1+\chi^{(1)}+\gamma
I_{0} S_{0})$. The constant $S_{0}$ is fixed by
\begin{equation}
\left(1-\frac{k_{1}^{2}}{k_{0}^{2}}\right)\mathcal{E}^{2}(0)+
\left(3+\frac{k_{1}^{2}}{k_{0}^{2}}\right)S_{0}-4=0.
\label{eq=bcPoy}
\end{equation}
Combining Eqs. (\ref{eq=epsPoy}) and (\ref{eq=bcPoy}), the
transmitance can be expressed in a similar fashion as the linear
Fabry-Perot resonator as
\begin{equation}\label{airy}
T=\frac{1}{1+F\sin^{2}\left(k_1 d\right)},
\end{equation}
where $F=k_0(1-k_1/k_0)^2/(4k_1)$. Carefully noting that Eq.
(\ref{airy}) is a transcendental expression since $k_{1}$ depends
on $S_{0}$.

\begin{figure}[t]
\includegraphics[width=8.5cm,height=5cm]{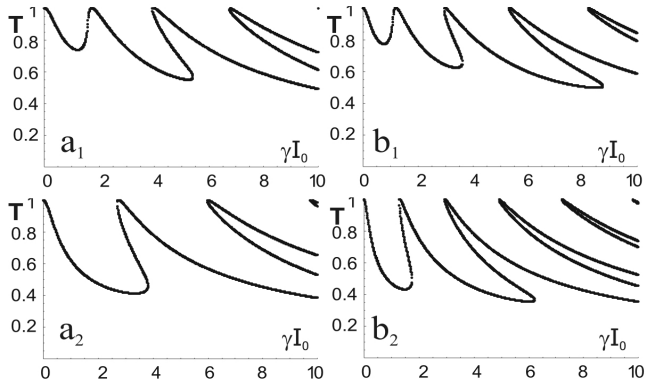}
\caption{Transmittance against nonlinear parameter for $(a_i)$
Poynting medium, $(b_i)$ Kerr medium with $k_{0}\:d=2
\pi$. For $i=1$, $\chi^{(1)}=1.25$; and $i=2$,
$\chi^{(1)}=5.25$.\\[-20pt]} \label{fig=curvas}
\end{figure}

Now, we compare the transmittance results of the Poynting and Kerr
media. The latter defined by
\begin{eqnarray}
\chi_{gen}&=&\chi^{(1)}+\gamma I_0 \mathcal{E}^2(z),\\[3pt]
\sigma_{gen}&=&0.
\end{eqnarray}

\begin{figure}[b]
\includegraphics[width=8.5cm,height=3cm]{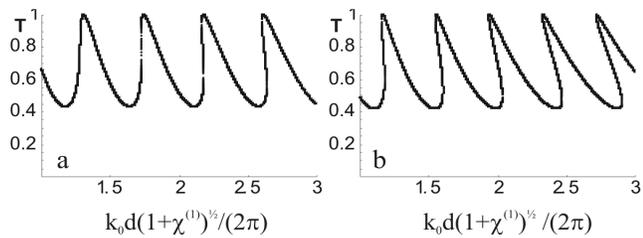}
\caption{Transmittance against dimensionless thickness for $(a)$
Poynting medium, $(b)$ Kerr medium with $\chi^{(1)}=5.25$ and
$\gamma I_{0}=2$.\\[-20pt]} \label{fig=curvas}
\end{figure}

The Kerr-NLFP transmittance results were taken from Ref.
\cite{mills1}. Figure 2 shows $T$ against the nonlinear parameter
$\gamma I_{0}$ for two different values of $\chi^{(1)}$. Figures
2($a_i$) and 2($b_i$) correspond to the Poynting and Kerr medium,
respectively. From these figures, it is apparent that the
transmittance of the Poynting-NLFP as well as the Kerr-NLFP are
multistable. However, for increasing values of $\chi^{(1)}$, the
peak transmittance separation diminishes for the Kerr medium while
it increases for the Poynting medium. Also, the Kerr
multistability appears for smallest values of the nonlinear
parameter $\gamma I_0$. The transmittance difference of both media
emphasizes the $I$ and $|E|^2$ nonequivalence. Figure 3 depicts
$T$ on the dimensionless thickness
$k_0d(1+\chi^{(1)})^{1/2}/(2\pi)$ enhancing the nonlinearity
difference of Poynting and Kerr media. Note that the departure
from Airy-type function for the Kerr medium is stronger than for
the Poynting medium.

On the other hand, Figure 4 shows that the Poynting nonlinear
susceptibility, $\chi_{gen}-\chi^{(1)}$, has a constant value
along the medium ($z$ coordinate). In return, the Kerr nonlinear
susceptibility varies periodically. This fact implies the
formation of a phase grating in the Kerr medium on the contrary to
the Poynting medium. Perhaps, this substantial difference can be
measured, and this could be the starting point to experimentally
identify a Poynting medium.

\subsection{Transformation properties
under spatial inversion and time reversal}

It is a fact that several unusual types of nonlinearities were
predicted before its experimental observation, as it was remarked,
for example, in the pioneer works of Baranova et al.
\cite{baranova}. With the aim to elucidate the isotropic medium
requirements to observe the new phenomena, those authors pointed
out the necessity of a analysis about transformation properties of
electromagnetic quantities under rotations, spatial inversion and
time reversal. Therefore, this symmetry analysis is also necessary
to delimit the Poynting medium requirements.
\begin{figure}[t]
\includegraphics[width=6cm,height=3.5cm]{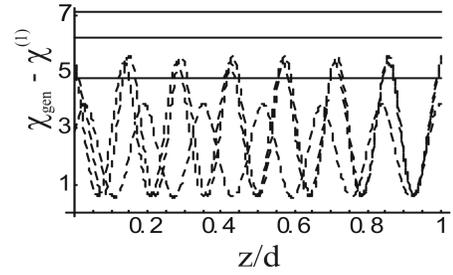}%
\caption{Nonlinear susceptibility against dimensionless spatial
coordinate for each of the three solutions compatible with the
boundary conditions. Continuous line: Poynting medium. Broken
line: Kerr medium. The parameter values are $\gamma I_0=9$,
$\chi^{(1)}=1.25$, and $k_0\:d$ as defined in Fig. 2.\\[-20pt]}
\label{fig=indices}
\end{figure}

The magnitude that characterizes the electromagnetic response of a
Poynting medium is their nonlinear susceptibility $\chi^{(P)}$
which is linear on the time-averaged Poynting vector
$\langle\textbf{S}\rangle$, as follows from the constitutive
relations [Eq. (\ref{constitutive})]. In a general form, it can be
written as
\begin{equation}
\chi^{(P)}_{ij}=\gamma_{ijk}
\:\big\langle\textbf{S}(\textbf{r},t)\big\rangle_k,\label{relconst}
\end{equation}
with $i,j,k = x,y,z$ and
\begin{equation}
\big\langle\textbf{S}(\textbf{r},t)\big\rangle_k=\frac{1}{T}
\int^{t+T}_t
 \big[\textbf{E}(\textbf{r},t')\times
 \textbf{H}(\textbf{r},t')\big]_k\:
dt',\label{chivss}
\end{equation}
where $\textbf{E}$ and $\textbf{H}$ are harmonics of period
$\tau=2\pi/\omega$ and time interval $T\gg \tau$. The
susceptibility tensor transforms as even under spacial inversion
$(r \rightarrow -r)$ and time reversal $(t \rightarrow -t)$,
contrary to the Poynting vector and its time-averaged which
transforms odd under spatial inversion and time reversal, i.e.
$\big\langle\textbf{S}(\textbf{r},t)\big\rangle\rightarrow
-\big\langle\textbf{S}(-\textbf{r},t)\big\rangle$ and
$\big\langle\textbf{S}(\textbf{r},t)\big\rangle\rightarrow
-\big\langle\textbf{S}(\textbf{r},-t)\big\rangle$, respectively
\cite{jackson}. Then, a medium possessing a linear connection
between $\chi_{ij}$ and $\langle S\rangle_z$ should be
noninvariant with respect to spatial inversion and time reversal.
Otherwise, the space-time symmetry will be violated in the
constitutive relation [Eq. (23)].

The lack of parity symmetry under inversion of coordinates is
proper of materials without inversion center, i. e. Poynting
nonlinearity should be only possible in
\textit{noncentrosymmetric} materials. There are several materials
candidates to possess a Poynting nonlinearity, as for example the
cubic crystals with zincblended structure such as GaAs, InSb and
others. In these materials intensity-dependent transmission and
bistability was experimentally observed \cite{InSb}. Also,
isotropic homogeneous liquids formed by nonracemic mixtures or
solutions of mirror-asymmetric (chiral) molecules with strong
nonlinear optical susceptibility, as product of several nonlinear
processes \cite{korotev}, are also feasible of posses a Poynting
nonlinearity. In addition, parity under time reversal should be
violated in Poynting media. This means that any weak dissipative
process, that converts field energy into heat, is necessary to
remove the rule relating to the $t\rightarrow -t$ transformation.
For example, either very weak absorbtion or current flow by
external quasi-static field, that basically do not affect the wave
propagation at light frequency $\omega$, would ensure the medium
non-invariance under time reversal.

We believe that, in spite of experimental works are required, the
above preliminary analysis could stimulate further discussions
regarding the existence of Poynting media.

\section{Conclusions}

In summary, we derive a new formalism in terms of dimensionless
variables related with the time-averaged Poynting vector and field
amplitude modulus within the Fabry-Perot framework. The
S-Formalism shows explicitly that the energy intensity and
squared-field amplitude modulus are only equivalents for a single
plane wave propagating in a linear infinite or semi-infinite
medium. Otherwise, they are nonequivalents. Besides, the
S-Formalism presents two important advantages: it permits to
directly monitor the time-averaged Poynting vector in the medium
and it avoids approximations, such as SVEA, simplification of the
boundary conditions, and so on. To emphasize this nonequivalence
we introduce the Poynting medium, whose nonlinearity is
proportional to the intensity instead of the electric
squared-field amplitude modulus such as in the Kerr medium. We
find marked disagreement in the transmittance of both media, which
support the differences between I and $|E|^2$. Also, a space-time
symmetry analysis shows that the Poynting nonlinearity should be
only possible in noncentrosymmetric materials.

The statements and analysis pointed out here constitute an advance
on theoretical views of basic concepts in electrodynamics. The
S-Formalism could be important in problems where the time-averaged
Poynting vector must be rigourously monitored like in
photoconductor or photorefractive materials. Further to this
particular case studied here, this new approach leaves open the
possibility of new physical results in actual topics on nonlinear
wave propagation such as spatial solitons, wave mixing and others.
Finally, we leave open the possibility that experimental
techniques, based on intensity dependent phase changes of a
Gaussian beam such as \textit{Z-Scan Technique} \cite{z-scan},
could not truly measure Kerr-type nonlinearity. On the contrary,
they could be measuring a Poynting-type instead.

\begin{acknowledgements} The authors thank Prof. Boris Ya. Zel'dovich and
anonymous Referee for the suggestions about symmetry properties of
Poynting media. The authors also thank Victor Waveluk for valuable
advices. A. L. thanks to CLAF-CNPq fellowship.
\end{acknowledgements}

\end{document}